\documentclass[pre,10pt,aps,longbibliography,nofootinbib]{revtex4-1}

 \usepackage{verbatim}
 \usepackage{amsmath}
 \usepackage{amssymb}
 \usepackage{amsthm}
 \usepackage{latexsym}
 \usepackage{amsfonts}
 \usepackage{epsfig}
 \usepackage{epstopdf}
 \usepackage{wasysym} 
 \usepackage{color}
 \definecolor{darkblue}{rgb}{0,0,.5}
 \usepackage[linktocpage, colorlinks=true, linkcolor=darkblue, citecolor=darkblue]{hyperref}
 \usepackage[all]{hypcap}
 \usepackage[makeroom]{cancel}

\newcommand{\C}[1]{{\cal{#1}}}

\newcommand{\lr}[1]{{\left\langle {#1}\right\rangle}}
\newcommand{\rl}[0]{{\rangle\langle}}

\begin{document}

\title{Entropy production as change in observational entropy}

\author{Philipp Strasberg}
\affiliation{F\'isica Te\`orica: Informaci\'o i Fen\`omens Qu\`antics, Departament de F\'isica, Universitat Aut\`onoma de Barcelona, 08193 Bellaterra (Barcelona), Spain}

\date{\today}

\begin{abstract}
 The thermodynamic entropy of the universe should increase in time by virtue of the second law of thermodynamics. Within 
 the open system paradigm -- where the universe is composed out of a small (quantum or classical) system coupled to one 
 or multiple heat baths -- the always positive entropy production is supposed to faithfully capture the increase in 
 thermodynamic entropy of the universe. Indeed, recent progress has shown how to derive under very mild assumptions an 
 always positive quantity, which resembles many features of entropy production. Unfortunately, this approach does not 
 express the entropy production as a change in thermodynamic entropy of the universe. That is to say, the very 
 definition of the basic thermodynamic state function measuring the entropy of the system-bath composite remained 
 unclear. 
 
 Here, we will put forward a different approach based on the recently introduced ``observational entropy'' by 
 \u{S}afr\'anek, Deutsch and Aguirre, which generalizes standard thermodynamic entropy to time-dependent 
 out-of-equilibrium processes in an isolated system~\cite{SafranekDeutschAguirrePRA2019a, SafranekDeutschAguirrePRA2019b, 
 SafranekDeutschAguirreArXiv2019}. We show that the observational entropy with respect to an arbitrary measurement of 
 the system and an energetic measurement of the bath fulfills the following key properties: 
 
 (i) Its change is always positive under even milder assumptions as used in the previous approach, (ii) it has a clear 
 information theoretic interpretation as the lack of knowledge about the exact microstate for a given energy 
 $E$ of the bath, (iii) in the ideal weak-coupling case the change in observational entropy of the bath can be shown to 
 be proportional to its change in energy divided by temperature. 
 
 Thus, the change in observational entropy provides a legitimate candidate to quantify entropy production, which 
 is now by construction expressed as a change in a thermodynamic state function. Furthermore, we show that the change 
 in observational entropy is typically smaller than the proposed entropy production definition of the previous approach. 
 However, in the weak coupling limit they are quantitatively almost identical. At the end, we also verify all our 
 general findings by using a `microcanonical master equation' derived in the Markovian, weak coupling limit. This 
 equation keeps track of energetic changes in the bath and therefore contains more information than typically used 
 weak-coupling master equations in quantum and stochastic thermodynamics. 
\end{abstract}

\maketitle

\newtheorem{mydef}{Definition}[section]
\newtheorem{lemma}{Lemma}[section]
\newtheorem{thm}{Theorem}[section]
\newtheorem{crllr}{Corollary}[section]
\newtheorem*{thm*}{Theorem}
\theoremstyle{remark}
\newtheorem{rmrk}{Remark}[section]

\section{Introduction}

\subsection{Motivation}

Entropy production is a central concept to quantify the irreversibility of nonequilibrium systems and dissipative 
structures and as such, it is of central importance to understand many situations encountered in biology, chemistry, 
physics and engineering~\cite{KondepudiPrigogineBook2007}. Within the open system paradigm, where a system exchanges 
energy and entropy with its environment, entropy production $\Sigma$ is typically expressed as 
\begin{equation}
 \Sigma = \Delta S_S + \Delta S_\text{env} \ge 0,
\end{equation}
Here, $\Delta S_S$ ($\Delta S_\text{env}$) measures the change in thermodynamic entropy of the system (environment). 
Positivity of entropy production and the second law of nonequilibrium thermodynamics become equivalent statements then. 
Furthermore, if the environment is composed out of several ideal heat baths $\nu$, the change in thermodynamic entropy 
of the environment can be expressed as $\Delta S_\text{env} \equiv \sum_\nu \Delta S_\nu$ with 
$\Delta S_\nu = -\beta_\nu Q_\nu$. Here, $Q_\nu$ is the heat flow from bath $\nu$ \emph{into} the system and the 
proportionality constant is the inverse temperature $\beta_\nu$ of the bath. Then, entropy production takes on the 
familiar form 
\begin{equation}\label{eq EP}
 \Sigma = \Delta S_S - \sum_\nu \beta_\nu Q_\nu \ge 0.
\end{equation}

In this paper we are interested in small systems $S$, which can be dominated by fluctuations and might show quantum 
effects. Also within this scenario the notion of entropy production plays an essential role, e.g., to quantify the 
efficiency of molecular motors or quantum heat engines in stochastic and quantum thermodynamics (see, e.g., 
Refs.~\cite{SeifertRPP2012, KosloffEntropy2013, SchallerBook2014} for introductions). Conventionally, positivity of 
entropy production is derived from an effective master or Fokker-Planck equation, which describes the dynamics of the 
system after tracing out the bath under various assumptions~\cite{SeifertRPP2012, KosloffEntropy2013, SchallerBook2014, 
BreuerPetruccioneBook2002}. While this reassures the thermodynamic consistency of the derived master or Fokker-Planck 
equation, it is somewhat unsatisfactory as it is not clear how entropy production emerges from the underlying 
microscopic (i.e., Hamiltonian) dynamics of the system and the baths. 

This paper claims to provide a valid and satisfactory microscopic derivation of entropy production for a large class 
of open system scenarios. Indeed, as we will review in the next section, recent progress in nonequilibrium statistical 
mechanics has almost provided a satisfactory answer to this question. But one important point was left open: so far 
it was impossible to find a definition for a thermodynamic state function, which, first, describes the entropy of the 
system-bath composite (the `universe') and, second, whose change is equal to the entropy production. In that 
respect previous derivations failed to show that entropy production measures the change in thermodynamic entropy of 
the universe. 

\subsection{Previous derivations of positivity of entropy production}
\label{sec previous results}

Deriving the laws of thermodynamics from an underlying microscopic (i.e., Hamiltonian) picture is a central theme of 
statistical mechanics since its beginning. Here, we briefly review the approach by Esposito, Lindenberg and Van den 
Broeck~\cite{EspositoLindenbergVandenBroeckNJP2010} where a microscopic (i.e., Hamiltonian) derivation of the second 
law of nonequilibrium thermodynamics for a driven system in contact with multiple heat baths was provided. In case of 
a single heat bath, this result was independently derived in Ref.~\cite{TakaraHasegawaDriebePLA2010}, see also 
Ref.~\cite{JarzynskiJSP1999} for an early derivation in the classical case where a system acts sequentially with 
different baths prepared at different temperatures. Similar work and extensions of this approach can be found in 
Refs.~\cite{HasegawaEtAlPLA2010, EspositoVandenBroeckEPL2011, DeffnerJarzynskiPRX2013, ReebWolfNJP2014, 
GooldPaternostroModiPRL2015, BarraSciRep2015, StrasbergEtAlPRX2017}. Furthermore, related work showed how to derive 
`dissipation inequalities' on a Hamiltonian basis for a system in contact with a single heat 
bath~\cite{JarzynskiPRL1997, JarzynskiPRE1997, JarzynskiJSP1999, KawaiParrondoVandenBroeckPRL2007, 
VaikuntanathanJarzynskiEPL2009, ParrondoVandenBroeckKawaiNJP2009}. In this approach it is assumed that the system 
starts in equilibrium and can relax back to it after the process has finished. While employing similar mathematical 
manipulations, this approach is, strictly speaking, less general than the one that will be reviewed here which takes 
explicitly the nonequilibrium nature of the initial and final system states into account. Also multiple baths are 
commonly not treated there, but see Ref.~\cite{ParrondoVandenBroeckKawaiNJP2009}. For completeness, let us also mention 
an alternative approach for cyclic Hamiltonian dynamics (i.e., where the initial and final Hamiltonian are the same) 
based on the notion of complete passivity~\cite{UzdinSaarPRX2018}. For the rest of this section we will focus on 
the derivation in Ref.~\cite{EspositoLindenbergVandenBroeckNJP2010}. We will also establish notation here and illustrate 
the problem with this approach, which we will eventually overcome later on. We remark that we choose a quantum 
mechanical notation throughout this paper, but the corresponding classical manipulations are analogous. 

Within the standard paradigm of open system theory, we will assume that the dynamics of the universe are modeled by 
the Hamiltonian $H_\text{tot}(\lambda_t) = H_S(\lambda_t) + H_{SB} + H_B$. Here, $H_S(\lambda_t)$ is the system 
Hamiltonian with $\lambda_t$ some externally specified driving protocol (e.g., a changing electric field). Furthermore, 
$H_B$ describes the bath Hamiltonian and $H_{SB}$ the interaction between system and bath. The present approach can be 
generalized to the case of multiple heat baths and a driven interaction Hamiltonian, but for ease of presentation 
we refrain from doing so in this section. Now, the only assumption we will add is that the initial state of the 
universe is given by 
\begin{equation}\label{eq initial state}
 \rho_\text{tot}(0) = \rho_S(0)\otimes \pi_B,
\end{equation}
where $\rho_S(0)$ is arbitary and $\pi_B \equiv e^{-\beta H_B}/\C Z_B$ 
with $\C Z_B = \mbox{tr}_B\{e^{-\beta H_B}\}$ denotes the canonical equilibrium state of the bath. This assumption 
is essential for the following and conventionally used in open system theory~\cite{BreuerPetruccioneBook2002}. 
An extension to correlated initial states is also possible for a single heat bath~\cite{SeifertPRL2016, 
StrasbergEspositoPRE2019}, see also Refs.~\cite{JarzynskiJSM2004, CampisiTalknerHaenggiPRL2009, JarzynskiPRX2017} for 
related `dissipation inequalities' and Ref.~\cite{UzdinSaarPRX2018} for cyclic Hamiltonian processes. In this paper, 
however, we will exclusively focus on initially decorrelated states. 

Now, let us denote by $S_\text{vN}(\rho) = -\mbox{tr}\{\rho\ln\rho\}$ the von Neumann entropy of the state $\rho$. Then, 
since the global dynamics are entropy-preserving, we find immediately with the help of Eq.~(\ref{eq initial state}) that 
\begin{equation}\label{eq marginal entropy change}
 \Delta S_\text{vN}[\rho_S(t)] + \Delta S_\text{vN}[\rho_B(t)] = I[\rho_{SB}(t)] \ge 0,
\end{equation}
where $\Delta S_\text{vN}[\rho(t)] = S_\text{vN}[\rho(t)] - S_\text{vN}[\rho(0)]$ denotes the change in von Neumann 
entropy and $I[\rho_{SB}(t)] = S_\text{vN}[\rho_S(t)] + S_\text{vN}[\rho_B(t)] - S_\text{vN}[\rho_{SB}(t)] \ge 0$ is the 
always positive mutual information. It is tempting to view Eq.~(\ref{eq marginal entropy change}) already as the 
entropy production: at least it is always positive and given by a change in a state function, namely the sum of the local 
von Neumann entropies. Indeed, if the `bath' itself is microscopically small, it makes sense to identify the heat flux 
directly via $-\beta Q = \Delta S_\text{vN}[\rho_B(t)]$~\cite{BeraEtAlNC2017}. Then, 
Eq.~(\ref{eq marginal entropy change}) formally takes on the conventional form of entropy production, compare with 
Eq.~(\ref{eq EP}) in case of a single heat bath. Unfortunately, for a mesoscopic or macrocopic heat bath this does not 
provide a satisfactory resolution as the mutual information can always be bounded by 
$I(\rho_{SB}) \le 2_Q\ln\dim(\C H_S)$, where $\C H_S$ denotes the Hilbert space of the system $S$, the factor $2_Q$ is 2 
for quantum systems and 1 for classical systems, and we assumed that (quite naturally) $\dim(\C H_S)\le\dim(\C H_B)$. 
Thus, e.g., for a two-level system the entropy production would be bounded from above \emph{for all times} by $2_Q\ln2$, 
which is clearly in general not the case. As a counterexample it suffices to consider, e.g., a driven system subjected 
to a laser field which dissipates energy into its environment. The entropy production in this case should rather scale 
extensively with time. This point was recently emphasized in Ref.~\cite{PtaszynskiEspositoArXiv2019}. 

Therefore, one employs a second important step by noting the exact identity 
$\Delta S_\text{vN}[\rho_B(t)] = \beta\Delta E_B - D[\rho_B(t)\|\pi_B]$, where 
$\Delta E_B = \mbox{tr}_B\{H_B[\rho_B(t) - \rho_B(0)]\}$ is the change in bath energy and 
$D[\rho\|\sigma] = \mbox{tr}\{\rho(\ln\rho-\ln\sigma)\} \ge 0$ is the always positive relative entropy. Then, 
one identfies 
\begin{equation}\label{eq EP open}
 \tilde\Sigma \equiv \Delta S_S + \beta\Delta E_B = I[\rho_{SB}(t)] + D[\rho_B(t)\|\pi_B] \ge 0
\end{equation}
as the entropy production, here denoted by $\tilde\Sigma$ to distinguish it from our approach put forward later on. 
Upon further identifying (minus) the change in bath energy with the heat flux into the system, $Q = -\Delta E_B$, one 
obtains $\tilde\Sigma \equiv \Delta S_S - \beta Q$ as in Eq.~(\ref{eq EP}) for a single heat bath. It should be noted, 
however, that the correct identification of heat is subtle outside the limit of a weakly coupled Markovian bath, 
compare, e.g., with the discussion in the classical case~\cite{SeifertPRL2016, TalknerHaenggiPRE2016, JarzynskiPRX2017, 
MillerAndersPRE2017, StrasbergEspositoPRE2017} or various approaches in quantum 
thermodynamics~\cite{EspositoOchoaGalperinPRB2015, StrasbergEtAlNJP2016, BruchEtAlPRB2016, NewmanMintertNazirPRE2017, 
BruchLewenkopfVonOppenPRL2018, StrasbergEtAlPRB2018, DouEtAlPRB2018, StrasbergEspositoPRE2019}. 
Therefore, we will mainly use the notation $\Delta E_B$ here. 

Let us summarize the picture so far: Eq.~(\ref{eq EP open}) proposes a definition of entropy production, which 
\begin{enumerate}
 \item[(i)] is positive for \emph{arbitary} system-bath dynamics and arbitary system and bath sizes based solely on 
 assumption~(\ref{eq initial state}), 
 \item[(ii)] has a natural information theoretic interpretation\footnote{Namely as deviation of the actual bath state 
 $\rho_B(t)$ from the equilibrium state $\pi_B$ as measured by the relative entropy. See also 
 Ref.~\cite{PtaszynskiEspositoArXiv2019} for a more specific discussion in that direction.}, and 
 \item[(iii)] has the conventional form as in phenomenological nonequilibrium dynamics [Eq.~(\ref{eq EP})], whereas 
 attention has to be paid to the point that the identification with heat is only valid in the weak coupling 
 limit.\footnote{In the weak coupling limit $\Delta E_B$ can be rigourously linked to the heat exchanges $Q$ (known, 
 e.g., from a weak coupling master equation) by use of the two point measurement approach, see
 Ref.~\cite{EspositoHarbolaMukamelRMP2009}.}  
\end{enumerate}

While these achievements are remarkable, they leave open the question what is the definition of thermodynamic 
entropy for the universe $S_\text{univ}$, whose change equals the entropy production. If $\tilde\Sigma$ were the 
correct definition of entropy production, then the thermodynamic entropy of the universe at time $t$ must be 
\begin{equation}
 \begin{split}
  S_\text{univ}(t) &= S_\text{vN}[\rho_S(t)] + \beta\mbox{tr}_B\{H_B\rho_B(t)\} + c \\
                   &= S_\text{vN}[\rho_S(t)] + S_\text{vN}[\rho_B(t)] + D[\rho_B(t)\|\pi_B] + \tilde c \\
                   &= S_\text{vN}[\rho_S(t)] + S_\text{vN}[\rho_B(t)] + D[\rho_B(t)\|\rho_B(0)] + \tilde c
 \end{split}
\end{equation}
such that $\Delta S_\text{univ}(t) = \tilde\Sigma$. Here $c$ and $\tilde c$ are arbitrary additive constants, which do 
not depend on time. Obviously, the first line seems to be a rather awkward definition for thermodynamic entropy. 
However, the second line reveals that this definition is identical to the local von Neumann entropies plus the relative 
entropy $D[\rho_B(t)\|\pi_B]$. While looking more reasonable, this definition depends on the initial Gibbs state of the 
bath (and therefore on the initial temperature), which is also unacceptable: the formal \emph{definition} of a 
thermodynamic entropy should not depend on such details. For the third line we noticed that $\pi_B = \rho_B(0)$ such 
that we can get rid of the formal dependence on $\pi_B$ in the definition, but unfortunately the so defined entropy 
would no longer be a state function. 

The goal of this paper is exactly to overcome this deficiency, while retaining the important properties (i) to 
(iii) above. This requires us to put forward a different approach to the problem. 

\subsection{Observational entropy}

Our notion of entropy for the universe, whose change we will identify with the entropy production, is based on the 
recently introduced notion of ``observational'' (or ``coarse grained'') entropy by \u{S}afr\'anek, Deutsch and 
Aguirre~\cite{SafranekDeutschAguirrePRA2019a, SafranekDeutschAguirrePRA2019b, SafranekDeutschAguirreArXiv2019}. 
They claim that observational entropy provides a satisfactory generalization of standard thermodynamic 
entropy to time-dependent out-of-equilibrium processes for arbitary isolated quantum and classical systems. What was 
not investigated in Ref.~\cite{SafranekDeutschAguirrePRA2019a, SafranekDeutschAguirrePRA2019b, 
SafranekDeutschAguirreArXiv2019}, however, is the question whether observational entropy also applies to the 
open system paradigm considered here (i.e., where the isolated `system' is divided into a small system of primary 
interest and the surrounding heat baths). 

The basic idea of observational entropy is relatively easy explained in the classical case. Suppose we perform a set of 
measurements on a thermodynamic system described by partitioning the phase space $\Gamma$ into different cells according 
to some coarse-graining procedure. Thus, each coarse graining $\C C_\alpha = \{\C C_{\alpha_i}\}_i$ partitions the phase 
space into non-overlapping regions corresponding to one measurement outcome, i.e., $\Gamma = \cup_i \C C_{\alpha_i}$ and 
$\C C_{\alpha_i}\cap\C C_{\alpha_j} = \emptyset$ for $i\neq j$. Here, $\alpha\in\{1,\dots,M\}$ denotes the different 
measurements (e.g., position of particles, energy, magnetization, etc.). The thermodynamic entropy of a system measured 
in such a way is then postulated to be 
\begin{equation}
 S_\text{obs} \equiv -\sum_{\C C_1,\dots,\C C_M} p_{\C C_1,\dots,\C C_M}\ln\frac{p_{\C C_1,\dots,\C C_M}}{V_{\C C_1,\dots,\C C_M}},
\end{equation}
where $p_{\C C_1,\dots,\C C_M}$ denotes the probability to obtain the joint measurement outcome $(\C C_1,\dots,\C C_M)$ 
and $V_{\C C_1,\dots,\C C_M}$ describes the number of possible microstates associated to it. 

A couple of remarks are in order. First, as also noted in Ref.~\cite{SafranekDeutschAguirrePRA2019a, 
SafranekDeutschAguirrePRA2019b, SafranekDeutschAguirreArXiv2019}, not every observational entropy is thermodynamically 
meaningful. Its usefulness depends crucially on the chosen observables. Second, for a `fine-grained' measurement of all 
positions and momenta of all particles, we reproduce the Gibbs-Shannon entropy. On the other hand, if we measure only the 
energy $E$, and if this is a conserved quantity, then we get the Boltzmann (surface but not volume!) entropy $\ln V_E$ 
back. Finally, a quantum extension by using projection operators in Hilbert space is possible but more complicated 
as the \emph{order} of the measurements now plays a role~\cite{SafranekDeutschAguirrePRA2019a, 
SafranekDeutschAguirrePRA2019b}. \u{S}afr\'anek \emph{et al.}~argue, however, that for typical situations encountered 
in thermodynamics, the effect of the non-commutativity of the measurements becomes very small. In our case we will 
actually choose commuting measurements, namely one measurement performed on the system Hilbert space and 
one on the bath Hilbert space. The construction of observational entropy is then straightforward, see below. 

\subsection{Outline}

The rest of this paper is organized as follows: In Sec.~\ref{sec general} we will present our choice of measurements, 
which defines the observational entropy. Then, we will show that the change of observational entropy provides a 
legitimate candidate for entropy production by verifying the points (i) to (iii) above for \emph{arbitary} open system 
dynamics as in Sec.~\ref{sec previous results}. We will also quantitaively compare our definition of entropy production 
with the previous approach and the case of multiple heat baths is treated in Sec.~\ref{sec multiple baths}. Additional 
generally valid observations are discussed in Sec.~\ref{sec add obs}. While the results in Sec.~\ref{sec general} are 
very general, they are necessarily also a bit abstract. Hence, in Sec.~\ref{sec MME} we will derive a `microcanonical 
master equation' for a system weakly coupled to a Markovian bath. This master equation takes explicitly into account 
changes of the bath energy. Within this important approximation we will then see that we are also able to verify all 
the above properties. Finally, we will conclude our findings in Sec.~\ref{sec conclusions}. 

\section{General picture}
\label{sec general}

Our definition of thermodynamic entropy for the system-bath setup requires two measurements: an arbitrary (but 
fine-grained) measurement of the system described by a set of rank-1 projectors $|s\rl s|$, 
$s\in\{1,\dots,\dim\C H_S\}$, and a measurement of the bath energy described by a set of projection operators 
$\Pi_{E,\delta} = \sum_{E_i\in(E-\delta,E]} \Pi_{E_i,0}$. Here, $\delta$ denotes a suitable width of the measured 
energy window (to be further specified in Sec.~\ref{sec ideal}) and $\Pi_{E_i,0}$ projects on a sharp energy $E_i$ such 
that $H_B \Pi_{E_i,0} = E_i\Pi_{E_i,0}$ (notice that $\Pi_{E_i,0}$ is not a rank-1 projector in case of exact 
degeneracies). For an arbitary system-bath state $\rho_{SB}$ the average post measurement state is given by 
\begin{equation}\label{eq CPTP meas map}
 \sum_{s,E} |s\rl s|\Pi_{E,\delta}\rho_{SB}\Pi_{E,\delta}|s\rl s| 
 = \sum_{s,E} p_{sE}|s\rl s|\otimes \rho_B(s,E).
\end{equation}
Here, $p_{sE} = \lr{s|\mbox{tr}_B\{\Pi_{E,\delta}\rho_{SB}\}|s}$ denotes the probability to obtain measurement outcome 
$(s,E)$ and the state of the bath conditioned on that outcome is 
$\rho_B(s,E) = \lr{s|\Pi_{E,\delta}\rho_{SB}\Pi_{E,\delta}|s}/p_{sE}$. Note that the average post measurement 
state~(\ref{eq CPTP meas map}) has lost all quantum correlations, but is in general classically correlated. Now, the 
observational entropy~\cite{SafranekDeutschAguirrePRA2019a, SafranekDeutschAguirrePRA2019b, 
SafranekDeutschAguirreArXiv2019} becomes in this case 
\begin{equation}
 \boxed{
  S_\text{obs} \equiv -\sum_{s,E} p_{sE} \ln\frac{p_{sE}}{V_{E,\delta}},
 }
\end{equation}
where $V_{E,\delta} = \mbox{tr}_B\{\Pi_{E,\delta}\}$ is the number of microstates in the bath with respect to a given 
energy window $(E-\delta,E]$. 

The claim of this paper is now that $\Sigma \equiv \Delta S_\text{obs}(t) = S_\text{obs}(t) - S_\text{obs}(0)$ is the 
entropy production of a system (perhaps subjected to a time-dependent driving $\lambda_t$) coupled to a single heat 
bath (multiple baths are treated in Sec.~\ref{sec multiple baths}) by verifying the points (i) to (iii) from above. 
As in Sec.~\ref{sec previous results} we will need an additional condition on the initial system-bath state, 
which is, however, milder than Eq.~(\ref{eq initial state}). Furthermore, we remark that our approach is in some sense 
close to the two-point measurement approach~\cite{EspositoHarbolaMukamelRMP2009} with the difference that we do not 
have to perfectly measure the energy of the bath, see Sec.~\ref{sec ideal}. 

Before verifying the points (i) to (iii) in Sec.~\ref{sec positivity EP}, we start with two general identities in 
Sec.~\ref{sec two identities} followed by a discussion in Sec.~\ref{sec ideal} of how small we have to choose the width 
$\delta$ in an experiment such that all theoretical claims of this paper remain true even if $\delta>0$. 

\subsection{Two general identities}
\label{sec two identities}

We start by noting that for any system-bath state $\rho_{SB}$ 
\begin{equation}\label{eq identity 1}
 S_\text{obs} = S_\text{Sh}(p_s) + S_\text{obs}^{E_B} - I(p_{sE}).
\end{equation}
Here, $S_\text{Sh}(p_s) \equiv -\sum_s p_s\ln p_s$ is the Shannon entropy of the probability distribution 
$p_s = \sum_E p_{sE}$ and 
\begin{equation}\label{eq obs entropy B}
 S_\text{obs}^{E_B} \equiv -\sum_E p_E \ln\frac{p_E}{V_{E,\delta}}
\end{equation}
can be interpreted as the observational entropy of the bath alone with respect to an energy measurement. Finally, 
$I(p_{sE}) = \sum_{s,E}p_{sE} \ln(p_{sE}/p_s p_E)$ is the classical mutual information between the measurement results 
of the system state and the bath energy. 

The second identity concerns only $S_\text{obs}^{E_B}$. For an arbitrary bath state $\rho_B$ let 
$\rho_B(E) \equiv \Pi_{E,\delta}\rho_B\Pi_{E,\delta}/p_E$ be the post-measurement state of the bath conditioned on 
outcome $E$ ignoring the measurement result $s$. This state is obtained with probability 
$p_E = \sum_s p_{sE} = \mbox{tr}_B\{\Pi_{E,\delta}\rho_B\}$. Furthermore, we introduce the microcanonical equilibrium 
state $\rho_\text{mic}(E) = \Pi_{E,\delta}/V_{E,\delta}$ with respect to a given energy $E$. Then, a straightforward 
calculation reveals 
\begin{equation}\label{eq identity 2}
 S_\text{obs}^{E_B} = S_\text{vN}\left[\sum_Ep_E\rho_B(E)\right] + \sum_E p_E D[\rho_B(E)\|\rho_\text{mic}(E)],
\end{equation}
where we used the identity 
\begin{equation}\label{eq theorem 1110}
 S_\text{Sh}(p_E) + \sum_i p_E S_\text{vN}[\rho_B(E)] = S_\text{vN}\left[\sum_E p_E\rho_B(E)\right],
\end{equation}
which holds since the states $\rho_B(E)$ are supported on orthogonal subspaces, see Theorem~11.10 in 
Ref.~\cite{NielsenChuangBook2000}. Equation~(\ref{eq identity 2}) tells us that the observational entropy of the bath 
is identical to the fine-grained (von Neumann) entropy of the average post-measurement state \emph{plus} the additional 
ignorance (measured by the relative entropy) due to not knowing the precise microstate of the bath for a given energy 
$E$. 

\subsection{Initial state and ideal measurement limit}
\label{sec ideal}

For the moment, let us consider the initial state~(\ref{eq initial state}), generalizations are discussed in 
Sec.~\ref{sec add obs}. The crucial ingredient to show positivity of the second law in our approach is that, initially 
at time $t=0$, the observational entropy of the bath $S_\text{obs}^{E_B}$ must coincide with the von Neumann entropy of 
the average post measurement state $\sum_E p_E(0)\rho_B(E,0)$. This means that, using Eq.~(\ref{eq identity 2}), the 
following expression must vanish: 
\begin{equation}\label{eq difference entropies}
 S_\text{obs}^{E_B}(0) - S_\text{vN}\left[\sum_E p_E(0)\rho_B(E,0)\right] 
 = \sum_E p_E(0) D[\rho_B(E,0)\|\rho_\text{mic}(E)]
\end{equation}
For a Gibbs state of the bath we have 
\begin{equation}
 p_E(0) = \pi_E \equiv \sum_{E_i\in(E-\delta,E]} \frac{e^{-\beta E_i} V_{E_i,0}}{\C Z_B}, ~~~ 
 \rho_B(E,0) = \pi_B(E) \equiv \sum_{E_i\in(E-\delta,E]} \Pi_{E_i,0} \frac{e^{-\beta E_i}}{p_E(0)\C Z_B}.
\end{equation}
Here, the projector $\Pi_{E_i,0}$ has rank $V_{E_i,0} = \mbox{tr}_B\{\Pi_{E_i,0}\}$ (which is greater than one in case 
of exact degeneracies). Using this, it becomes clear that Eq.~(\ref{eq difference entropies}) can be written as 
\begin{equation}
 S_\text{obs}^{E_B}(0) - S_\text{vN}(\pi_B) 
 = \sum_E \sum_{E_i\in(E-\delta,E]} \frac{e^{-\beta E_i}V_{E_i,0}}{Z_B}\ln\frac{e^{-\beta E_i}V_{E,\delta}}{\sum_{E_j\in(E-\delta,E]} e^{-\beta E_j} V_{E_j,0}},
\end{equation}
which vanishes in the ideal theoretical limit $\delta\rightarrow0$. For finite $\delta$ we proceed by looking at 
the argument of the logarithm, 
\begin{equation}
 \frac{e^{-\beta E_i}V_{E,\delta}}{\sum_{E_i\in(E-\delta,E]} V_{E_i,0} e^{-\beta E_i}} 
 = \frac{e^{\beta\delta_i}V_{E,\delta}}{\sum_{j\in[0,\delta)} e^{\beta\delta_j}V_{E-\delta_j,0}},
\end{equation}
where we defined $E_j = E-\delta_j$. Now, if the bath is macroscopically large, we expect that we can replace 
the sums by integrals and by using the mean value theorem for integration, we end up with 
\begin{equation}
 \frac{e^{\beta\delta_i}V_{E,\delta}}{\int_0^\delta dx e^{\beta x}V_{E-x,0}} 
 = \frac{e^{\beta\delta_i}V_{E,\delta}}{e^{\beta\xi}\int_0^\delta dx V_{E-x,0}} 
 = e^{\beta(\delta_i-\xi)} = 1 + \beta(\delta_i-\xi) + \C O[(\beta\delta)^2]
\end{equation}
with $\xi\in[0,\delta)$ and we used $\int_0^\delta dx V_{E-x,0} = V_{E,\delta}$. Thus, in the limit 
\begin{equation}\label{eq condition experiment}
 \boxed{
  \beta\delta \ll 1
 }
\end{equation}
the observational entropy practically coincides with the von Neumann entropy for a thermal state. Experimentally, 
this condition has to be met in order to ensure positivity of entropy production if the bath is initially in a Gibbs 
state. In the following we will assume that the energy window $\delta$ is chosen small enough such that 
Eq.~(\ref{eq condition experiment}) holds and we will henceforth simply denote $V_E = V_{E,\delta}$ and 
$\Pi_E = \Pi_{E,\delta}$. 

\subsection{Verifying points (i) to (iii) and comparison with the previous approach for a single heat bath}
\label{sec positivity EP}

For an arbitrary initial system-bath state $\rho_{SB}(0)$ the change in observational entropy can be expressed as 
\begin{equation}
 \begin{split}
  \Delta S_\text{obs} =&~ \Delta S_\text{Sh}[p_s(t)] + \Delta S_\text{obs}^{E_B} - \Delta I[p_{sE}(t)] \\
  =&~ S_\text{Sh}[p_s(t)] - S_\text{Sh}[p_s(0)] 
  + S_\text{vN}\left[\sum_Ep_E(t)\rho_B(E,t)\right] - S_\text{vN}\left[\sum_Ep_E(0)\rho_B(E,0)\right] \\
  & + \sum_E p_E(t) D[\rho_B(E,t)\|\rho_\text{mic}(E)] - \sum_E p_E(0) D[\rho_B(E,0)\|\rho_\text{mic}(E)] 
  - I_{s:E}(t) + I_{s:E}(0),
 \end{split}
\end{equation}
where we used Eqs.~(\ref{eq identity 1}) and~(\ref{eq identity 2}). Here, $p_s(t) = \lr{s|\rho_S(t)|s}$ and 
$\rho_B(E,t) = \Pi_E\rho_B(t)\Pi_E/p_E(t)$. Next, we assume the initial state and width $\delta$ to be as described in 
Sec.~\ref{sec ideal}, such that $S_\text{obs}^{E_B}(0) = S_\text{vN}(\pi_B)$. This allows us to confirm  
\begin{equation}
 \begin{split}
  \Delta S_\text{obs} =&~ S_\text{Sh}[p_s(t)] - S_\text{Sh}[p_s(0)] 
  + S_\text{vN}\left[\sum_Ep_E(t)\rho_B(E,t)\right] - S_\text{vN}(\pi_B) \\
  & + \sum_E p_E(t) D[\rho_B(E,t)\|\rho_\text{mic}(E)] - I[p_{sE}(t)].
 \end{split}
\end{equation}
Now, as the entropy is preserved during any unitary evolution, we get 
$S_\text{Sh}[p_s(0)] + S_\text{vN}(\pi_B) = S_\text{vN}[\rho_{SB}(t)]$ where $\rho_{SB}(t)$ is the 
time-evolved state starting from the initial state $\rho_{SB}(0) = \sum_s p_s(0)|s\rl s|\otimes\pi_B$. Writing also 
$S_\text{vN}[\rho_{SB}(t)] = S_\text{vN}[\rho_S(t)] + S_\text{vN}[\rho_B(t)] - I[\rho_{SB}(t)]$, we get 
\begin{equation}
 \begin{split}\label{eq Delta S obs 1}
  \Delta S_\text{obs} =&~ S_\text{Sh}[p_s(t)] - S_\text{vN}[\rho_S(t)] 
  + S_\text{vN}\left[\sum_Ep_E(t)\rho_B(E,t)\right] - S_\text{vN}[\rho_B(t)] \\
  & + \sum_E p_E(t) D[\rho_B(E,t)\|\rho_\text{mic}(E)] + I[\rho_{SB}(t)] - I[p_{sE}(t)].
 \end{split}
\end{equation}
The positivity of $\Delta S_\text{obs}$ is now evident. First, by using that a projective measurement increases the 
entropy on average, Theorem~11.9 in Ref.~\cite{NielsenChuangBook2000}, we confirm that 
\begin{equation}
 S_\text{Sh}[p_s(t)] \ge S_\text{vN}[\rho_S(t)], ~~~ 
 S_\text{vN}\left[\sum_Ep_E(t)\rho_B(E,t)\right] \ge S_\text{vN}[\rho_B(t)].
\end{equation}
We expect both contributions, however, to be rather small. First, the change in system entropy due to the measurement 
is at most $\ln\dim(\C H_S)$ and will likely be much smaller. Especially, we are free to choose the basis of the final 
system measurement at time $t$ such that we can let it coincide with the eigenbasis of $\rho_S(t)$. Second, a large 
change in bath entropy due to the final measurement requires the existence of large coherences 
$\mbox{tr}_B\{\Pi_E\rho_B(t)\Pi_{E'}\} \neq 0$ between different energy sectors $E\neq E'$. This also seems very 
unlikely as it would imply the existence of macrosopic Schr\"odinger cat states in the bath. 
Finally, another small contribution comes from the difference in mutual information, which always obeys 
\begin{equation}
 \begin{split}\label{eq inequality mutual info}
  I[\rho_{SB}(t)] &= D[\rho_{SB}(t)\|\rho_S(t)\otimes\rho_B(t)] \\
  &\ge D\left[\sum_{s,E} p_{sE}(t) |s\rl s|\otimes\rho_\text{mic}(E)\left\|\sum_s p_s(t)|s\rl s|\otimes\sum_E p_E(t)\rho_\text{mic}(E)\right]\right. = I[p_{sE}(t)],
 \end{split}
\end{equation}
This follows from monotonicity of relative entropy~\cite{UhlmannCMP1977, OhyaPetzBook1993} by noting that 
$\sum_{s,E} p_{sE}(t) |s\rl s|\otimes\rho_\text{mic}(E) = \Phi\rho_{SB}(t)$ with the completely positive and 
trace-preserving map defined via 
\begin{equation}
 \Phi\rho_{SB} \equiv 
 \sum_E \mbox{tr}_B\left\{|s\rl s|\Pi_E\rho_{SB}|s\rl s|\Pi_E\right\} \otimes \rho_\text{mic}(E).
\end{equation}

Thus, all together we can conclude that 
\begin{equation}\label{eq second law interpretation}
 \boxed{
  \Delta S_\text{obs} \gtrsim \sum_E p_E(t) D[\rho_B(E,t)\|\rho_\text{mic}(E)] \ge 0,
 }
\end{equation}
where we used the symbol ``$\gtrsim$'' to indicate that the difference between the two sites of the inequality is 
expected to be rather small in the typical situation of a large heat bath. The only part, which can scale extensively 
with time, is $\sum_E p_E(t) D[\rho_B(E,t)\|\rho_\text{mic}(E)]$. Therefore, we conclude that \emph{entropy production 
arises because for a given energy $E$ of the bath we lose track of its exact microstate} compared to the maximal 
uninformative microcanonical ensemble, where all microstates are assumed to be equally likely. Thus, we have confirmed 
the points (i) and (ii) from above. 

Next, to confirm (iii), we want to show that for a weakly coupled, macroscopic bath the change in observational bath 
entropy is proportional to its change in energy. For that purpose it is indeed crucial to assume that the bath is 
initially in a Gibbs state. Then, we write $p_E(t) = \pi_E + \epsilon q_E(t)$ with $\pi_E = e^{-\beta E}V_E/\C Z_B$ 
and $q_E(t)$ is a set of numbers such that $\sum_E q_E(t) = 0$. Now, our assumption is that $\epsilon$ is a small 
parameter, i.e., the \emph{distribution of energies} in the bath remains close to the canonical probabilities 
throughout the time-evolution. This should be typically justified for a weakly coupled, macroscopic bath. Notice that 
this does \emph{not} imply that the entire bath state $\rho_B(t)$ is close to the Gibbs state $\pi_B$. Then, we can 
write 
\begin{equation}
 \Delta S_\text{obs}^{E_B} 
 = -\sum_E [\pi_E + \epsilon q_E(t)]\ln\frac{\pi_E + \epsilon q_E(t)}{V_E} + \sum_E \pi_E \ln\frac{\pi_E}{V_E}
 =\beta\Delta E_B + \C O(\epsilon^2)
\end{equation}
with the chang ein bath energy $\Delta E_B = \epsilon\sum_E E q_E(t)$. Hence, 
\begin{equation}
 \boxed{
  \Delta S_\text{obs} = \Delta S_\text{Sh}[p_S(t)] + \Delta S_\text{obs}^{E_B} - I[p_{sE}(t)] 
  \approx \Delta S_\text{Sh}[p_S(t)] - \beta Q \ge 0,
 }
\end{equation}
where we ignored the small contribution $I[p_{sE}(t)]$ at the end. We have also identified $Q = -\Delta E_B$, which is 
justified in the limit considered here to derive $\Delta S_\text{obs}^{E_B} \approx \beta\Delta E_B$. 

Thus, as a preliminary conclusion, we have shown that $\Sigma \equiv \Delta S_\text{obs}$ fulfills the three desired 
properties (i) to (iii) and therefore, provides a more suitable candidate for entropy production then $\tilde\Sigma$ 
from Eq.~(\ref{eq EP open}) because the latter cannot be expressed as the change of a meaningful thermodynamic entropy 
for the system and the bath. 

Nevertheless, it is instructive to compare $\Sigma$ and $\tilde\Sigma$ quantitatively. 
From Eqs.~(\ref{eq Delta S obs 1}) and~(\ref{eq EP open}) we obtain 
\begin{equation}
 \begin{split}\label{eq comparison help}
  \Sigma - \tilde\Sigma = &~ S_\text{Sh}[p_s(t)] + S_\text{vN}\left[\sum_Ep_E(t)\rho_B(E,t)\right] 
  - S_\text{vN}[\rho_{SB}(t)] + \sum_E p_E(t) D[\rho_B(E,t)\|\rho_\text{mic}(E)] - I[p_{sE}(t)] \\ 
  &- I[\rho_{SB}(t)] - D[\rho_B(t)\|\pi_B].
 \end{split}
\end{equation}
We consider first the very last term. This becomes after a little massage 
\begin{equation}
 \begin{split}
  D[\rho_B(t)\|\pi_B] 
  &= \mbox{tr}_B\left\{\rho_B(t)\left[\ln\rho_B(t) - \sum_E \Pi_E \ln\rho_\text{mic}(E)\rho_B(E)\right]\right\} \\
  &= -S_\text{vN}[\rho_B(t)] - \sum_E p_E(t)\mbox{tr}_B\{\rho_B(E,t)\ln\rho_\text{mic}(E)\} - \sum_E p_E(t)\ln\pi_B(E).
 \end{split}
\end{equation}
By combining this result with the second and forth term of Eq.~(\ref{eq comparison help}), we verify 
\begin{equation}
 S_\text{vN}\left[\sum_Ep_E(t)\rho_B(E,t)\right] + \sum_E p_E(t) D[\rho_B(E,t)\|\rho_\text{mic}(E)] - D[\rho_B(t)\|\pi_B] = S_\text{vN}[\rho_B(t)] - D[p_E(t)\|\pi_B(E)],
\end{equation}
where we also made use of Eq.~(\ref{eq theorem 1110}). Thus, we end up with the compact expression 
\begin{equation}\label{eq comparison EPs 0}
 \Sigma - \tilde\Sigma = S_\text{Sh}[p_s(t)] - S_\text{vN}[\rho_S(t)] - D[p_E(t)\|\pi_B(E)] - I[p_{sE}(t)].
\end{equation}
Taken together, the first two terms are non-negative (and identical to zero for classical systems), whereas each of 
the remaining two terms is negative. In the weak coupling regime investigated above, we expect all terms to be rather 
small. In fact, if $p_E(t) = \pi_E + \epsilon q_E(t)$ as above, then $D[p_E(t)\|\pi_B(E)] = \C O(\epsilon^2)$ such that 
we can conclude $\Sigma\approx\tilde\Sigma$. Outside this regime, the only possibly unbounded term is 
$D[p_E(t)\|\pi_B(E)]$ as the number of populated energy levels for a very small $\delta$ can become very large. We 
therefore expect that, typically, we have 
\begin{equation}\label{eq comparison EPs}
 \boxed{
  \tilde\Sigma \gtrsim \Sigma \ge 0.
 }
\end{equation}
This means that the previous approach typically yields a larger entropy production than our novel definition. 
Intuitively, this makes sense: in our approach we ideally know the entire distribution of energies in the bath whereas 
in the previous approach only knowing the average energy flow to the bath is sufficient. Clearly, our approach stores 
more information and hence, less entropy is produced. 

\subsection{Extension to multiple heat baths}
\label{sec multiple baths}

The extension to multiple heat baths labeled by $\nu\in\{1,\dots,N\}$ is straightforward by measuring the energy of 
each bath. The observational entropy is in this case 
\begin{equation}
 S_\text{obs} = -\sum_{s,E_1,\dots,E_N} p_{sE_1\dots E_N}\ln\frac{p_{sE_1\dots E_N}}{V_{E_1}\dots V_{E_N}}.
\end{equation}
Here, we used that the number of microstates naturally factorizes, 
$V_{E_1\dots E_N} = \mbox{tr}_{B_1\dots B_N}\{\Pi_{E_1}\dots\Pi_{E_N}\} = V_{E_1}\dots V_{E_N}$, where $\Pi_{E_\nu}$ 
describes the projector associated to measurement outcome $E_\nu$ of bath $\nu$. Furthermore, the natural 
generalization of the initial state~(\ref{eq initial state}) to multiple baths is 
\begin{equation}\label{eq initial state multiple baths}
 \rho_\text{tot}(0) = \rho_S(0)\otimes\pi_{B_1}\otimes\dots\otimes\pi_{B_N},
\end{equation}
which was also used in Ref.~\cite{EspositoLindenbergVandenBroeckNJP2010}. We note that every bath can have initially 
a different inverse temperature, i.e., $\pi_{B_\nu} = e^{-\beta_\nu H_B^{(\nu)}}/\C Z_{B_\nu}$. 

Under these circumstances (assuming that $\delta$ is chosen as in Sec.~\ref{sec ideal}) we easily confirm that the 
initial observational entropy is identical to 
\begin{equation}
 S_\text{obs}(0) = S_\text{vN}\left[\sum_s p_s(0)|s\rl s|\otimes\pi_{B_1}\otimes\dots\otimes\pi_{B_N}\right] 
 = S_\text{vN}[\rho_\text{tot}(t)].
\end{equation}
Furthermore, the observational entropy at time $t$ can be split into its `local' parts and its correlations, similar 
to Eq.~(\ref{eq identity 1}). Specifically, 
\begin{equation}
 \begin{split}
  S_\text{obs}(t) 
  &= -\sum_s p_s(t)\ln p_s(t) - \sum_\nu \sum_{E_\nu} p_{E_\nu}(t)\ln\frac{p_{E_\nu}(t)}{V_{E_\nu}}
 - \sum_{s,E_1,\dots,E_N} p_{sE_1\dots E_N}(t) \ln\frac{p_{sE_1\dots E_N}(t)}{p_s(t)p_{E_1}(t)\dots p_{E_N}(t)} \\
 &\equiv S_\text{Sh}[p_s(t)] + \sum_\nu S_\text{obs}^{E_{B_\nu}}(t) - I_\text{cor}[p_{sE_1\dots E_N}(t)].
 \end{split}
\end{equation}
Further use of relation~(\ref{eq identity 2}) reveals that 
\begin{equation}
 S_\text{obs}(t) 
 = S_\text{Sh}[p_S(t)] + \sum_\nu S_\text{vN}\left[\sum_{E_\nu} p_{E_\nu}\rho_{B_\nu}(E_\nu,t)\right] + 
 \sum_\nu \sum_{E_\nu} p_{E_\nu} D[\rho_{B_\nu}(E_\nu,t)\|\rho_\text{mic}(E_\nu)] - I_\text{cor}[p_{sE_1\dots E_N}(t)].
\end{equation}

Thus, similarly to Eq.~(\ref{eq Delta S obs 1}), the change in observational entropy can be split into a family of 
terms, whose non-negativity is evident: 
\begin{equation}
 \begin{split}
  \Delta S_\text{obs} =&~ S_\text{Sh}[p_S(t)] - S_\text{vN}[\rho_S(t)] 
  + \sum_\nu\left\{S_\text{vN}\left[\sum_{E_\nu} p_{E_\nu}\rho_{B_\nu}(E_\nu,t)\right] - S_\text{vN}[\rho_{B_\nu}(t)]\right\} \\
  & + \sum_\nu \sum_{E_\nu} p_{E_\nu} D[\rho_{B_\nu}(E_\nu,t)\|\rho_\text{mic}(E_\nu)] 
  + I_\text{cor}[\rho_\text{tot}(t)] - I_\text{cor}[p_{sE_1\dots E_N}(t)].
 \end{split}
\end{equation}
As before, we expect the non-negativity of the first line to be rather small (especially, it is exactly zero for 
classical systems) such that 
\begin{equation}
 \Delta S_\text{obs} \gtrsim \sum_\nu \sum_{E_\nu} p_{E_\nu} D[\rho_{B_\nu}(E_\nu,t)\|\rho_\text{mic}(E_\nu)] 
 + I_\text{cor}[\rho_\text{tot}(t)] - I_\text{cor}[p_{sE_1\dots E_N}(t)].
\end{equation}
Here, we introduced the notation 
\begin{equation}
 I_\text{cor}[\rho_\text{tot}(t)] 
 \equiv S_\text{vN}[\rho_S(t)] + \sum_\nu S_\text{vN}[\rho_{B_\nu}(t)] -S_\text{vN}[\rho_\text{tot}(t)] 
 = D[\rho_\text{tot}(t)\|\rho_S(t)\otimes\rho_{B_1}(t)\otimes\dots\otimes\rho_{B_N}(t)].
\end{equation}
Since we can also write 
\begin{equation}
 \begin{split}
  I_\text{cor}&[p_{sE_1\dots E_N}(t)] = \\
  &D\left[\sum_{s,E_1,\dots,E_N}p_{sE_1\dots E_N}(t)|s\rl s|\rho_\text{mic}(E_1)\dots\rho_\text{mic}(E_N)\left\|\sum_sp_s(t)|s\rl s|\sum_{E_1}p_{E_1}(t)\rho_\text{mic}(E_1)\dots\sum_{E_N}p_{E_N}(t)\rho_\text{mic}(E_N)\right]\right.,
 \end{split}
\end{equation}
we can confirm $I_\text{cor}[\rho_\text{tot}(t)] - I_\text{cor}[p_{sE_1\dots E_N}(t)] \ge 0$ similar to 
Eq.~(\ref{eq inequality mutual info}). However, in contrast to the case of a single heat bath, we can no longer expect 
this contribution to be small as the different baths can become correlated. This is especially true outside the 
weak coupling case. Therefore, in the case of multiple baths we expect that there are two major contributions to the 
entropy production. In any case, $\Delta S_\text{obs} \ge 0$ is ensured. 

Furthermore, following the same procedure as above, we can confirm in the weak coupling case that 
$\Delta S_\text{obs}^{E_\nu} = \beta_\nu\Delta E_{B_\nu}$. In addition, we expect in the weak coupling limit that 
$I_\text{cor}[\rho_\text{tot}(t)] \gtrsim I_\text{cor}[p_{sE_1\dots E_N}(t)] \approx 
\sum_\nu I[p_{sE_\nu}(t)]$, i.e., each bath acts like a separate bath entering independently the master equation 
describing the system~\cite{EspositoHarbolaMukamelRMP2009, KosloffEntropy2013, SchallerBook2014}.\footnote{A critical 
discussion of this point can be found in Ref.~\cite{MitchisonPlenioNJP2018}.} Hence, we can write 
$\Delta S_\text{obs}(t) \approx \Delta S_\text{Sh}[p_s(t)] - \sum_\nu\beta_\nu Q_\nu \ge 0$. This provides a 
microscopic derivation of the phenomenological second law of nonequilibrium thermodynamics~(\ref{eq EP}). 

\subsection{Additional observations}
\label{sec add obs}

We end this general section with a couple of interesting observations: 

\emph{Observation 1.} The crucial ingredient to prove point~(i), positivity of $\Delta S_\text{obs}$, is that 
Eq.~(\ref{eq difference entropies}) vanishes. This is not only true for a Gibbs state and a small enough measurement 
width $\delta$. Indeed, Eq.~(\ref{eq difference entropies}) can be zero for many different initial energy 
distributions $p_E(0)$ as long as the distribution of microstates within a given energy window is very close to the 
microcanonical ensemble. The initial state of the bath can even contain quantum coherences between different energy 
sectors as those get killed during the initial measurement. 

\emph{Observation 2.} One can also choose different measurements of the bath and positivity of $\Delta S_\text{obs}$ 
will still hold as long as the initial observational entropy of the bath coincides with the von Neumann entropy of the 
average post measurement state. In that respect the energy only seems to be an outstanding observable due to its 
connection to the first law. To capture the effect of multiple conserved quantities, we can consider additional 
measurements, e.g., of the energy and particle number of the bath in case of a grand-canonical reservoir. 

\emph{Observation 3.} We also do not expect the initial product state assumption to be crucial. An initially correlated 
system-bath state lowers the entropy production by at most $I[p_{sE}(0)]$, which is typically negligible with respect 
to the positive terms appearing in Eq.~(\ref{eq second law interpretation}). 

\emph{Observation 4.} Instead of taking into account correlations between the measurement results of the system state 
and the bath energy, we could also neglect them in the definition of observational entropy. All the three points~(i) 
to~(iii) would remain valid for the choice $\tilde S_\text{obs}(t) \equiv S_\text{Sh}[p_s(t)] + S_\text{obs}^{E_B}(t)$. 
In particular, we would typically have $\Delta S_\text{obs} \approx \Delta\tilde S_\text{obs}$. 

\emph{Observation 5.} Finally, we emphasize that none of our results depends on the particular form of the Hamiltonian. 
Especially, the system-bath coupling could be time-dependent, $H_{SB} = H_{SB}(\lambda_t)$, and even the bath 
Hamiltonian could depend on time, $H_B = H_B(\lambda_t)$. 

\section{The microcanonical master equation}
\label{sec MME}

In this section we illustrate our general findings in the limit of a weakly coupled, Markovian bath. In contrast to 
conventional master equations~\cite{SeifertRPP2012, KosloffEntropy2013, SchallerBook2014, BreuerPetruccioneBook2002}, 
we will derive a master equation describing the evolution of the system state \emph{and} the bath energies by using a 
correlated projection-operator method. Such a master equation was first derived by Esposito and 
Gaspard~\cite{EspositoGaspardPRE2003} and we repeat a (slightly more generalized) derivation in 
Appendix~\ref{sec MME derivation}. Here, we will investigate in detail the analytical and thermodynamic properties of 
this master equation, which was not done in Ref.~\cite{EspositoGaspardPRE2003}. We will call this approach the 
`microcanonical master equation' (MME) in the following. 

The MME is a Pauli-like rate master equation for the probabilities $p_{sE}(t)$ to find the system in state $s$ and 
the energy of the bath at $E$ at time $t$. It reads 
\begin{equation}\label{eq Pauli MME}
 \partial_t p_{sE}(t) = \sum_{\alpha,\gamma}\sum_{s'} 
 \frac{2\pi}{V_E}\Re\{S_\alpha^{ss'}S_\gamma^{s's}f_{\alpha\gamma}(E,E+\epsilon_s-\epsilon_{s'})\} 
 \left[\frac{V_E}{V_{E+\epsilon_s-\epsilon_{s'}}}p_{s',E+\epsilon_s-\epsilon_{s'}}(t) - p_{sE}(t)\right].
\end{equation}
Here, the overall timescale of the dynamics is governed by the rate 
$\Re\{S_\alpha^{ss'}S_\gamma^{s's}f_{\alpha\gamma}(E,E+\epsilon_s-\epsilon_{s'})\}$. If one assumes that the 
system-bath coupling Hamiltonian reads $H_{SB} = \sum_\alpha S_\alpha\otimes B_\alpha$, where $S_\alpha$ ($B_\alpha$) 
are Hermitian system (bath) operators, then $S^{ss'}_\alpha \equiv \lr{s|S_\alpha|s'}$ describes the transition 
matrix elements with respect to the basis $|s\rangle$, which is assumed in this section to be the (non-degenerate) 
energy eigenbasis of $H_S$. Furthermore, the function 
$f_{\alpha\gamma}(E,E') \equiv \mbox{tr}_B\{\Pi_E B_\alpha\Pi_{E'} B_\gamma\}$ describes how well the energies in 
the bath get redistributed. In obeys the useful relations~(\ref{eq symmetries f}). 

We remark that Eq.~(\ref{eq Pauli MME}) reduces to Eq.~(42) of Ref.~\cite{EspositoGaspardPRE2003} in the limit of a 
single system coupling operator $S_\alpha = \delta_{\alpha,1}S$. Multiple baths can be easily included by summing over 
$\nu$ and adding this superscript to $S_\alpha$, $f$ and $V$, but we will only consider a single heat bath here. As 
this equation describes the time-evolution of all energies, we will indeed find out below that the dynamics of this 
equation are entropy dominated. Furthermore, we remark that driven system energies can be considered by replacing 
$\epsilon_s$ by $\epsilon_s(\lambda_t)$ provided that the change of energies is slow compared to the decay of the 
bath correlation functions. 

\subsection{Properties}
\label{sec MME properties}

\subsubsection{Reduction to the conventional Pauli master equation}
\label{sec Pauli BMS}

As a simple crosscheck we investigate the limit in which our MME reduces to the conventional Pauli master equation 
derived within the Born-Markov-secular approximation~\cite{BreuerPetruccioneBook2002}. Formally, we can write 
Eq.~(\ref{eq Pauli MME}) after summing over $E$ as 
\begin{equation}
 \partial_t p_s(t) = \sum_E\sum_{\alpha,\gamma}\sum_{s'} 
 \frac{2\pi}{V_E}\Re\{S_\alpha^{ss'}S_\gamma^{s's}f_{\alpha\gamma}(E,E+\epsilon_s-\epsilon_{s'})\} 
 \left[\frac{V_E}{V_{E+\epsilon_s-\epsilon_{s'}}}p_{E+\epsilon_s-\epsilon_{s'}|s'}(t)p_{s'}(t) - p_{E|s}(t)p_s(t)\right].
\end{equation}
Here, we have introduced the conditional probability $p_{E|s}(t) \equiv p_{sE}(t)/p_s(t)$ and we will now assume that 
this is approximately given by $p_{E|s}(t) \approx V_E e^{-\beta E}/\C Z_B$ for all $s$ and all times $t$. This 
simplifies the expression to 
\begin{equation}
 \partial_t p_s(t) = \frac{2\pi}{Z_B}\sum_E\sum_{\alpha,\gamma}\sum_{s'} 
 \Re\{S_\alpha^{ss'}S_\gamma^{s's}f_{\alpha\gamma}(E,E+\epsilon_s-\epsilon_{s'})\} 
 \left[e^{-\beta(E+\epsilon_s-\epsilon_{s'})}p_{s'}(t) - e^{-\beta E}p_s(t)\right].
\end{equation}
We now introduce the functions 
$g_{\alpha\gamma}(\epsilon_s-\epsilon_{s'}) \equiv \sum_E f_{\alpha\gamma}(E,E+\epsilon_s-\epsilon_{s'}) \frac{e^{-\beta E}}{\C Z_B}$, which obey the symmetries 
$g_{\alpha\gamma}(\epsilon_s-\epsilon_{s'}) = e^{\beta(\epsilon_s-\epsilon_{s'})} g_{\gamma\alpha}(\epsilon_{s'}-\epsilon_s)$ and $g_{\alpha\gamma}^*(\epsilon_s-\epsilon_{s'}) = g_{\gamma\alpha}(\epsilon_s-\epsilon_{s'})$. 
They allows us to write 
\begin{equation}\label{eq Pauli BMS}
 \partial_t p_s(t) = 2\pi\sum_{\alpha,\gamma}\sum_{s'} 
 \Re\{S_\alpha^{ss'}S_\gamma^{s's}g_{\alpha\gamma}(\epsilon_s-\epsilon_{s'})\} 
 \left[e^{\beta(\epsilon_{s'}-\epsilon_s)}p_{s'}(t) - p_s(t)\right].
\end{equation}
This corresponds to the typical Pauli master equation~\cite{BreuerPetruccioneBook2002} with the rates 
satisfying local detailed balance, i.e., the rate to jump from $s'$ to $s$ is enhanced by a factor 
$e^{\beta(\epsilon_{s'}-\epsilon_s)}$ compared to the inverse jump rate from $s$ to $s'$ if 
$\epsilon_{s'} > \epsilon_s$. 

\subsubsection{Conservation of energy}

To confirm conservation of energy, we note that 
\begin{equation}
 \begin{split}\label{eq conservation energy help}
  \sum_{s,E}\sum_{\alpha,\gamma}\sum_{s'} & (\epsilon_s+E) 
  \Re\{S_\alpha^{ss'}S_\gamma^{s's}f_{\alpha\gamma}(E,E+\epsilon_s-\epsilon_{s'})\} 
  \frac{1}{V_{E+\epsilon_s-\epsilon_{s'}}}p_{s',E+\epsilon_s-\epsilon_{s'}}(t) \\
  &= \sum_{s,E'}\sum_{\alpha,\gamma}\sum_{s'} (\epsilon_{s'}+E') 
  \Re\{S_\alpha^{ss'}S_\gamma^{s's}f_{\alpha\gamma}(E'-\epsilon_s+\epsilon_{s'},E')\} 
  \frac{1}{V_{E'}}p_{s'E'}(t) \\
  &= \sum_{s,E}\sum_{\alpha,\gamma}\sum_{s'} (\epsilon_s+E) 
  \Re\{S_\alpha^{ss'}S_\gamma^{s's}f_{\alpha\gamma}(E,E+\epsilon_s-\epsilon_{s'})\} 
  \frac{1}{V_{E}}p_{sE}(t),
 \end{split}
\end{equation}
where we used the symmetry relation~(\ref{eq symmetries f}) and made use of the freedom to relabel indices 
within the summation. Equation~(\ref{eq conservation energy help}) holds for any fixed time and even in 
presence of driving when $\epsilon_s = \epsilon_s(\lambda_t)$. In this case one confirms that 
\begin{equation}\label{eq conservation law}
 \frac{d}{dt}[\Delta E_S(t) + \Delta E_B(t)] \equiv \frac{d}{dt}\sum_{s,E}[\epsilon_s(\lambda_t)+E] p_{sE}(t) 
 = \sum_s \dot\epsilon_s(\lambda_t) p_s(t) \equiv \dot W. 
\end{equation}
This is the first law of thermodynamics in presence of driving. 

Interestingly, for an undriven system Eq.~(\ref{eq conservation energy help}) even implies that 
\begin{equation}
 \frac{d}{dt}\lr{f(\epsilon_s+E)} \equiv \frac{d}{dt}\sum_{s,E} f(\epsilon_s+E) p_{sE}(t) = 0
\end{equation}
for an arbitrary function $f(E_\text{tot})$ of the total energy $E_\text{tot} = \epsilon_s + E$. We can call this 
\emph{strict energy conservation}. It essentially implies that there is only one random variable in the problem 
(and not the two $\epsilon_s$ and $E$) because the distribution for $E_\text{tot}$ remains fixed for all times. 
This conclusion holds, however, only in absence of driving. 

\subsubsection{Steady state of the Pauli MME}
\label{sec MME steady state}

We here consider the case where $\epsilon_s$ is held fixed in time (i.e., $\dot\lambda_t = 0$) and we ask for which 
state $\bar p_{sE}$ the Pauli MME~(\ref{eq Pauli MME}) evaluates to zero. We call $\bar p_{sE}$ a steady state in this 
case. One point we can immediately recognize from Eq.~(\ref{eq Pauli MME}) is that every state $\bar p_{sE}$ which fulfills 
\begin{equation}
 \frac{V_E}{V_{E+\epsilon_s-\epsilon_{s'}}} = \frac{\bar p(s,E)}{\bar p(s',E+\epsilon_s-\epsilon_{s'})}
\end{equation}
is a steady state. However, due to the fact that the energy is strictly conserved, there are infinitely many possible 
steady states; indeed even infinitely many for every initial energy $\lr{\epsilon_s+E}(0) = E_0$ depending on how 
the probabilities are initially distributed. For instance, one possible steady state is the overall Gibbs state 
\begin{equation}\label{eq Gibbs SB}
 \bar p_{sE} = \frac{e^{-\beta\epsilon_s}}{\C Z_S} \frac{V_E e^{-\beta E}}{\C Z_B} \equiv \pi_{sE},
\end{equation}
where $\beta$ must be fixed through $E_0 = -\partial_\beta\ln(\C Z_S\C Z_B)$. 

On the other hand, imagine that we start with a definite initial condition such as 
$p_{sE}(t=0) = \delta_{s,0}\delta_{E,E_0}$ and we assume that the energy eigenvalues are ordered according to 
$\epsilon_n > \dots > \epsilon_1 > \epsilon_0 \equiv 0$. Then, the dynamics are restricted to the following states 
with energies $(\epsilon_s,E)$: 
\begin{equation}
 (0,E_0), (\epsilon_1,E_0-\epsilon_1), \dots, (\epsilon_{n-1},E_0-\epsilon_{n-1}), (\epsilon_n,E_0-\epsilon_n).
\end{equation}
Note that the dynamics is not restricted to jumps between nearest neighbours as one might be tempted to 
think.\footnote{The precise topology of the network depends strongly on the prefactor 
$\Re\{S_\alpha^{ss'}S_\gamma^{s's}f_{\alpha\gamma}(E,E+\epsilon_s-\epsilon_{s'})\}$. } 
The ratio of the rates to jump from one state to another are 
\begin{equation}\label{eq ratio rates MME}
 \frac{\text{rate}[(\epsilon_s,E_0-\epsilon_s)\rightarrow(\epsilon_{s'},E_0-\epsilon_{s'})]}{\text{rate}[(\epsilon_{s'},E_0-\epsilon_{s'})\rightarrow(\epsilon_s,E_0-\epsilon_s)]} = \frac{V_{E_0-\epsilon_{s'}}}{V_{E_0-\epsilon_s}},
\end{equation}
which can be interpreted as a purely entropic factor. Thus, the \emph{dynamics of the MME are entropy dominated}. 
Typically, one expects that $V_{E'} > V_E$ if $E'>E$. Then, the system tends to prefer low energies in order to 
increase the entropy of the environment. A particularly interesting case arises if the bath behaves like an ideal heat 
bath. Using Boltzmann's entropy formula, we infer that $V_E = e^{S_B(E)/k_B}$ where $S_B(E)$ is the entropy of the bath 
at energy $E$. Now, the assumption of an ideal heat bath enters by invoking the standard definition of temperature, 
$T^{-1} = S'_B(E)$, which allows us to derive 
$\frac{V_{E_0-\epsilon_{s'}}}{V_{E_0-\epsilon_s}} = e^{\beta(\epsilon_s-\epsilon_{s'})}$. This implies that the ratio 
of the rates~(\ref{eq ratio rates MME}) fulfills the conventional local detailed balance relation. One steady state of 
the MME with initial condition $p_{sE}(t=0) = \delta_{s,0}\delta_{E,E_0}$ is then given by 
\begin{equation}\label{eq steady state microcanonical}
 \bar p_{sE} = 
 \left\{\begin{array}{ll}
         \frac{e^{-\beta\epsilon_s}}{Z_S} \frac{1}{n+1} & \text{if } E = E_0 - \epsilon_s \\
         0 & \text{otherwise} \\
        \end{array}\right.
\end{equation}
That is to say, the system equilbrated to the canonical ensemble with the temperature imposed by the initial energy of 
the bath which becomes equally distributed over the available phase space. This is nothing else than the equivalence of 
ensembles, i.e., the reduced state of a weakly coupled subsystem is a canonical distribution if the entire system has 
a fixed energy $E_0$. Note that the above state indeed fulfills $\sum_{s,E} (\epsilon_s+E)\bar p_{sE} = E_0$. 

\subsection{Entropy production and observational entropy}
\label{sec ent prod obs ent}

Conventionally, the entropy production for the Pauli master equation~(\ref{eq Pauli BMS}) can be expressed as 
\begin{equation}\label{eq EP Pauli BMS}
 \dot{\tilde\Sigma}(t) = \frac{d}{dt} S_\text{Sh}[p_s(t)] - \beta\dot Q(t) 
 = -\left.\frac{\partial}{\partial t}\right|_{\lambda_t}D[p_s(t)\|\pi_s(\lambda_t)] \ge 0,
\end{equation}
where $\pi_s(\lambda_t) = e^{-\beta\epsilon_s(\lambda_t)}/\C Z_S(\lambda_t)$ is the instantaneous canonical 
equilibrium state of the system at time $t$. The derivative in Eq.~(\ref{eq EP Pauli BMS}) is evaluated with respect 
to a fixed $\lambda_t$ and positivity of the entropy production follows from the two facts that the dynamics are 
Markovian and that $\pi_s(\lambda_t)$ is an instantaneous steady state of the dynamics, see, e.g., 
Refs.~\cite{BreuerPetruccioneBook2002, KosloffEntropy2013, StrasbergEspositoPRE2019}. Finally, 
$\dot Q(t) = \sum_s \epsilon_s(\lambda_t)\partial_t p_s(t)$ is the heat flow into the system. 

Similarly, also for the MME we can derive an always positive entropy production rate by considering 
\begin{equation}\label{eq EP Pauli MME general}
 \dot\Sigma(t) \equiv -\left.\frac{\partial}{\partial t}\right|_{\lambda_t}D[p_{sE}(t)\|\bar p_{sE}(\lambda_t)] \ge 0,
\end{equation}
where $\bar p_{sE}(\lambda_t)$ is \emph{any} admissible steady state of the Pauli MME (independent of the initial 
condition), which is allowed to depend parametrically on time through $\lambda_t$. As we have a multitude of possible 
steady states, see Sec.~\ref{sec MME steady state}, there are many different possible choices, each leading to a 
different positive `entropy production' rate. The choice, which leads to the desired final result, turns out be the 
Gibbs state from Eq.~(\ref{eq Gibbs SB}), where $\epsilon_s = \epsilon_s(\lambda_t)$ is allowed to be time-dependent. 
Indeed, upon integration of Eq.~(\ref{eq EP Pauli MME general}) we have 
\begin{equation}
 \begin{split}
  \Sigma(t) &= 
  \int_0^t ds\left(-\frac{d}{ds} + \dot\lambda_s\frac{\partial}{\partial\lambda_s}\right) D[p_{sE}(s)\|\pi_{sE}(\lambda_s)] \\
  &= -D[p_{sE}(t)\|\pi_{sE}(\lambda_t)] + D[p_{sE}(0)\|\pi_{sE}(\lambda_0)] 
  - \int_0^t ds\dot\lambda_s\frac{\partial}{\partial\lambda_s} \sum_{s,E} p_{sE}(s)\ln\pi_{sE}(\lambda_s),
 \end{split}
\end{equation}
where we used the chain rule 
$\frac{d}{dt} = \left.\frac{\partial}{\partial t}\right|_{\lambda_t} + \dot\lambda_t\frac{\partial}{\partial\lambda_t}$. 
The particular form of $\pi_{sE}(\lambda_t)$ reveals after some straightforward manipulations that 
\begin{equation}
 \begin{split}
  \Sigma(t) =& -\sum_{s,E} p_{sE}(t)\ln\frac{p_{sE}(t)}{V_E} + \sum_{s,E} p_{sE}(0)\ln\frac{p_{sE}(0)}{V_E} \\
  &+ \beta\sum_{s,E} \left\{- [\epsilon_s(\lambda_t)+E] p_{sE}(t)  + [\epsilon_s(\lambda_0)+E] p_{sE}(0) + 
  \int_0^t ds\epsilon_s(\lambda_s)p_{sE}(s)\right\} \\
  =&~ \Delta S_\text{obs}.
 \end{split}
\end{equation}
Here, for the final step we used the first law~(\ref{eq conservation law}). Thus, we have confirmed that 
$\Sigma(t) = \Delta S_\text{obs} \ge 0$ also follows directly from the Pauli MME. 

Finally, we compare $\Sigma(t)$ with $\tilde\Sigma(t)$ obtained by integrating Eq.~(\ref{eq EP Pauli BMS}). 
First, we use that we can express the heat flow within the MME approach as
\begin{equation}
 -\beta Q(t) = \beta \sum_E E [p_E(t)-p_E(0)] 
 = \sum_E [p_E(t)- p_E(0)]\ln V_E + \Delta S_\text{Sh}[p_E(t)] + D[p_E(t)\|\pi_E],
\end{equation}
if we assume that $p_E(0) = \pi_E$. If the initial state $p_{sE}(0) = p_s(0)p_E(0)$ is furthermore decorrelated, 
we can confirm that 
\begin{equation}
 \Sigma(t) - \tilde\Sigma(t) = -I[p_{sE}(t)] - D[p_E(t)\|\pi_E] \le 0.
\end{equation}
This result is in direct analogy to our general finding~(\ref{eq comparison EPs 0}). Thus, we have re-derived all our 
general findings within the particularly imporant limit of a system weakly coupled to a Markovian bath. Notice that 
we could even derive a stronger statement in that limit, namely that the \emph{rate} of entropy 
production~(\ref{eq EP Pauli MME general}) is positive. This is usually not the case within the general setup of 
Sec.~\ref{sec general}. Conditions which ensure the positivity of the entropy production rate 
$\frac{d}{dt}S_\text{obs}(t)$ are missing; within the standard approach reviewed in Sec.~\ref{sec previous results} 
answers were partially found in Refs.~\cite{StrasbergEspositoPRE2019, StrasbergTBP}. 

\section{Conclusions}
\label{sec conclusions}

We have put forward a novel approach to understand and quantify entropy production in open (quantum) systems driven 
arbitrarily far from equilibrium. For this purpose we constructed a suitable notion of entropy for the entire universe 
(system plus bath) based on the recently introduced observational entropy from 
Refs.~\cite{SafranekDeutschAguirrePRA2019a, SafranekDeutschAguirrePRA2019b, SafranekDeutschAguirreArXiv2019}. 
Then, using very similar steps as in previous approaches (conservation of global von Neumann entropy, special -- but 
slightly more general -- form of the initial system-bath state, see Sec.~\ref{sec ideal}), we showed that the change 
in observational entropy is always positive for arbitrary dynamics, has a clear information-theoretic interpretation, 
and can be linked to the standard expression~(\ref{eq EP}) in the limit of weak system-bath coupling. Therefore, our 
novel notion fulfills the three minimum requirements (i) to (iii), but -- moreover and most importantly -- it fulfills 
them expressed as a change in a thermodynamically meaningful definition of global entropy. Thus, we were able to 
microscopically derive the statement that \emph{entropy production measures the change in thermodynamic entropy of the 
universe, which is always positive.} 

Quite interestingly, in the conventionally considered weak coupling limit we showed that the quantitative difference 
between our and the former approach is negligible. This reassures the consistency of both, our and the former approach. 
Outside the weak coupling regime, interesting difference could appear. As we have also stated the condition which needs 
to be fulfilled to test this theory experimentally [Eq.~(\ref{eq condition experiment})], we have the hope that it will 
be possible to measure the global entropy production in the future, for instance, in cold 
atoms~\cite{LewensteinSanperaAhufingerBook2012} or in electronic nanostructures coupled to mesoscopic heat 
baths~\cite{PekolaNatPhys2015}. 

Finally, we believe that our results provide strong evidence that observational entropy as advertised in 
Refs.~\cite{SafranekDeutschAguirrePRA2019a, SafranekDeutschAguirrePRA2019b, SafranekDeutschAguirreArXiv2019} provides 
a good candidate for thermodynamic entropy of isolated out-of-equilibrium systems. 

\subsection*{Acknowledgments}

I am grateful to Andreu Riera-Campeny and Andreas Winter for thoughtful comments. Also various stimulating discussions 
with Massimiliano Esposito about the nature of entropy production over the years are acknowledged. I am financially 
supported by the DFG (project STR 1505/2-1). I also acknowledge funding from the Spanish MINECO FIS2016-80681-P 
(AEI-FEDER, UE). 


\bibliography{/home/philipp/Documents/references/books,/home/philipp/Documents/references/open_systems,/home/philipp/Documents/references/thermo,/home/philipp/Documents/references/info_thermo,/home/philipp/Documents/references/general_QM,/home/philipp/Documents/references/math_phys,/home/philipp/Documents/references/equilibration}

\appendix
\section{Derivation of Eq.~(\ref{eq Pauli MME})}
\label{sec MME derivation}

We start with the standard system-bath Hamiltonian $H_\text{tot} = H_S + H_{SB} + H_B$ as usual and assume that the 
interaction can be decomposed as $H_{SB} = \sum_\alpha S_\alpha\otimes B_\alpha$, where $S_\alpha$ ($B_\alpha$) are 
Hermitian system (bath) operators. The standard time-convolutionless Nakajima-Zwanzig projection operator method 
predicts within the weak coupling and the interaction picture that~\cite{BreuerPetruccioneBook2002} 
\begin{equation}\label{eq Redfield ME}
 \partial_t\C P\tilde\rho(t) = \int_0^t ds \C P\C L_I(t)\C L_I(s)\C P\tilde\rho(t),
\end{equation}
where $\C L_I(t)\rho \equiv -i[H_{SB},\rho]$. As our projection superoperator we choose 
\begin{equation}
 \C P\rho = \sum_E \mbox{tr}_B\{\Pi_E\rho\} \otimes\rho_\text{mic}(E).
\end{equation}
Below, we will denote $\rho_S(E,t) \equiv \mbox{tr}_B\{\C P\rho(t)\} = \mbox{tr}_B\{\Pi_E\rho\}$. We remark that the 
validity of Eq.~(\ref{eq Redfield ME}) is only ensured if $(1-\C P)\rho(0) = 0$. Furthermore, we 
assume that $\mbox{tr}\{B_\alpha\rho_\text{mic}(E)\} = 0$ for all $\alpha$ and $\nu$. The latter step can be done 
without loss of generality~\cite{EspositoGaspardPRE2003}. 

Using $\Pi_E\Pi_{E'} = \delta_{E,E'}\Pi_E$, the trace over the bath degrees of freedom of Eq.~(\ref{eq Redfield ME}) 
yields 
\begin{equation}
 \begin{split}
  \sum_E \partial_t\tilde\rho_S(E,t) =& 
  -\sum_E\sum_{\alpha,\gamma}\int_0^t ds\left[\lr{B_\alpha(t)B_\gamma(s)}_E S_\alpha(t)S_\gamma(s)\tilde\rho_S(E,t) + 
  \lr{B_\gamma(s)B_\alpha(t)}_E\tilde\rho_S(E,t)S_\gamma(s)S_\alpha(t)\right] \\
  & +\sum_{E,E'}\sum_{\alpha,\gamma}\int_0^t ds 
  \left[\lr{B_\gamma(s)\Pi_E B_\alpha(t)}_{E'} S_\alpha(t)\tilde\rho_S(E',t)S_\gamma(s) + 
  \lr{B_\alpha(t)\Pi_E B_\gamma(s)}_{E'}S_\gamma(s)\tilde\rho_S(E',t)S_\alpha(t)\right].
 \end{split}
\end{equation}
Here, we defined the microcanonical average $\lr{\dots}_E \equiv \mbox{tr}_B\{\dots\rho_\text{mic}(E)\}$. To obtain 
an equation for $\rho_S(E,t)$, we drop the sum over $E$, which appears on all sides. Then, after coming back to the 
Schr\"odinger picture and after a change of integration variables $\tau = t-s$, we are left with 
\begin{equation}
 \begin{split}
  \partial_t\rho_S(E,t) =& -i[H_S,\rho_S(E,t)] 
  -\sum_{\alpha,\gamma}\int_0^t d\tau\left[\lr{B_\alpha(\tau)B_\gamma}_E S_\alpha S_\gamma(-\tau)\rho_S(E,t) + 
  \lr{B_\gamma(-\tau)B_\alpha)}_E\rho_S(E,t)S_\gamma(-\tau)S_\alpha\right] \\
  & +\sum_{E'}\sum_{\alpha,\gamma}\int_0^t d\tau 
  \left[\lr{B_\gamma(-\tau)\Pi_E B_\alpha}_{E'} S_\alpha\rho_S(E',t)S_\gamma(-\tau) + 
  \lr{B_\alpha(\tau)\Pi_E B_\gamma}_{E'}S_\gamma(-\tau)\rho_S(E',t)S_\alpha\right].
 \end{split}
\end{equation}
This is the microcanonical Redfield master equation, which only relies on the Born and weak-coupling approximation. 
As such, it is still quite general, but hard to deal with in practise and theory. 

We next take a look at the correlation functions. Denoting the energy of a single bath eigenstate by 
$E_i = E + \delta_i$, we obtain and approximate 
\begin{equation}
 \begin{split}
  \lr{B_\alpha(\tau)B_\gamma} 
  &= \sum_{E'} \frac{e^{i(E-E')\tau}}{V_E} \sum_{E_i\in(E-\delta,E]} \sum_{E'_i\in(E'-\delta,E']} 
  \mbox{tr}_B\{\Pi_{E_i,0}B_\alpha \Pi_{E'_i,0}|B_\gamma\} e^{i(\delta_i-\delta_{i'})\tau} \\
  &\approx \sum_{E'} \frac{e^{i(E-E')\tau}}{V_E} \mbox{tr}_B\{\Pi_E B_\alpha \Pi_{E'}B_\gamma\} 
  \equiv \sum_{E'} \frac{e^{i(E-E')\tau}}{V_E} f_{\alpha\gamma}(E,E').
 \end{split}
\end{equation}
This approximation is similar but not identical to the one of Sec.~\ref{sec ideal}. We expect it to be valid in the 
limit where $\delta$ is small compared to differences in the eigenspectrum of $H_S$ and if the bath correlation 
functions are peaked around $\tau=0$, i.e., in the limit typically associated with Markovianity. For later purposes we 
also note the symmetries 
\begin{equation}\label{eq symmetries f}
 f_{\alpha\gamma}(E,E') = f_{\gamma\alpha}(E',E), ~~~ f_{\alpha\gamma}^*(E,E') = f_{\gamma\alpha}(E,E') \in \mathbb{C}.
\end{equation}

In this Markovian limit we then send the integration limit $t$ to infinity and we are left with 
\begin{equation}
 \begin{split}
  \partial_t\rho_S(E,t) =& -i[H_S,\rho_S(E,t)] \\
  &- \sum_{E'}\sum_{\alpha,\gamma}\sum_{s,s'} S_\gamma^{ss'} \int_0^\infty d\tau 
  e^{i(E-E'-\epsilon_s+\epsilon_{s'})\tau} f_{\alpha\gamma}(E,E')
  \left\{\frac{1}{V_E} S_\alpha |s\rl s'|\rho_S(E,t) - \frac{1}{V_{E'}} S_\alpha\rho_S(E',t)|s\rl s'|\right\} \\
  &- \sum_{E'}\sum_{\alpha,\gamma}\sum_{s,s'} S_\gamma^{ss'} \int_0^\infty d\tau 
  e^{i(E'-E-\epsilon_s+\epsilon_{s'})\tau} f_{\gamma\alpha}(E,E')
  \left\{\frac{1}{V_E} \rho_S(E,t)|s\rl s'|S_\alpha - \frac{1}{V_{E'}}|s\rl s'|\rho_S(E',t)S_\alpha\right\},
 \end{split}
\end{equation}
where we decomposed $S_\gamma(\tau) = \sum_{s,s'} S_\gamma^{ss'} e^{i(\epsilon_s-\epsilon_{s'})\tau}|s\rl s'|$ in 
the (assumed to be non-degenerate) energy eigenbasis of $H_S$. Next, we use 
$\int_0^\infty dt e^{ixt} = \pi\delta(x)$, where we neglected any imaginary (Lamb shift) contributions. This allows 
us to write 
\begin{equation}
 \begin{split}
  \partial_t\rho_S(E,t) =& -i[H_S,\rho_S(E,t)] \\
  & -\sum_{\alpha,\gamma}\sum_{s,s'} \frac{\pi S_\gamma^{ss'}}{V_E} f_{\alpha\gamma}(E,E-\epsilon_s+\epsilon_{s'}) 
  \left\{S_\alpha|s\rl s'|\rho_S(E,t) - \frac{V_E}{V_{E-\epsilon_s+\epsilon_{s'}}} S_\alpha\rho_S(E-\epsilon_s+\epsilon_{s'},t)|s\rl s'|\right\} \\
  & -\sum_{\alpha,\gamma}\sum_{s,s'} \frac{\pi S_\gamma^{ss'}}{V_E} f_{\gamma\alpha}(E,E+\epsilon_s-\epsilon_{s'}) 
  \left\{\rho_S(E,t)|s\rl s'|S_\alpha - \frac{V_E}{V_{E+\epsilon_s-\epsilon_{s'}}} |s\rl s'|\rho_S(E+\epsilon_s-\epsilon_{s'},t)S_\alpha\right\}.
 \end{split}
\end{equation}

Finally, we apply the secular approximation assuming also that the differences in the energy spectrum of $H_S$ are 
non-degenerate. This amounts to replacing 
\begin{equation}
 \begin{split}
  \sum_{s,s'} S_\gamma^{ss'}S_\alpha|s\rl s'|\rho \mapsto \sum_{s,s'} S_\gamma^{ss'}S_\alpha^{s's}|s'\rl s'|\rho,~~~ 
  & \sum_{s,s'} S_\gamma^{ss'}S_\alpha\rho|s\rl s'| \mapsto \sum_{s,s'} S_\gamma^{ss'}S_\alpha^{s's}|s'\rl s|\rho|s\rl s'|, \\
  \sum_{s,s'} S_\gamma^{ss'}\rho|s\rl s'|S_\alpha \mapsto \sum_{s,s'} S_\gamma^{ss'}S_\alpha^{s's}\rho|s\rl s|,~~~ 
  & \sum_{s,s'} S_\gamma^{ss'}|s\rl s'|\rho S_\alpha \mapsto \sum_{s,s'} S_\gamma^{ss'}S_\alpha^{s's}|s\rl s'|\rho|s'\rl s|.
 \end{split}
\end{equation}
Hence, we end up with the microcanonical Born-Markov secular master equation: 
\begin{equation}
 \begin{split}
  \partial_t\rho_S(E,t) =& -i[H_S,\rho_S(E,t)] \\
  & -\sum_{\alpha,\gamma}\sum_{s,s'} \frac{\pi S_\gamma^{ss'}S_\alpha^{s's}}{V_E} f_{\alpha\gamma}(E,E-\epsilon_s+\epsilon_{s'}) 
  \left\{|s'\rl s'|\rho_S(E,t) - \frac{V_E}{V_{E-\epsilon_s+\epsilon_{s'}}} |s'\rl s|\rho_S(E-\epsilon_s+\epsilon_{s'},t)|s\rl s'|\right\} \\
  & -\sum_{\alpha,\gamma}\sum_{s,s'} \frac{\pi S_\gamma^{ss'}S_\alpha^{s's}}{V_E} f_{\gamma\alpha}(E,E+\epsilon_s-\epsilon_{s'}) 
  \left\{\rho_S(E,t)|s\rl s| - \frac{V_E}{V_{E+\epsilon_s-\epsilon_{s'}}} |s\rl s'|\rho_S(E+\epsilon_s-\epsilon_{s'},t)|s'\rl s|\right\}.
 \end{split}
\end{equation}
It is well-known that all quantum coherences exponentially die out with time and that the dynamics of this MME is 
well-captured by a Pauli-like master equation. Defining $p_{sE}(t) \equiv \lr{s|\rho_S(E,t)|s}$, we end up with 
the Pauli MME~(\ref{eq Pauli MME}) as stated in the main text.

\end{document}